\documentclass[11pt,preprintnumbers,pra,aps]{revtex4}
\usepackage{amssymb,amsmath,amsbsy,amsfonts,amsthm,mathtools,bm}
\usepackage[breaklinks,colorlinks,citecolor=blue,urlcolor=violet]{hyperref}
\usepackage[mediumqspace,mediumspace,amssymb,Gray]{SIunits}
\usepackage{dcolumn,longtable,rotating,tabularx,array,lscape,braket,url}

\usepackage{xcolor,soul,cancel,enumitem,ulem,verbatim,graphicx,epsfig}
\usepackage[T1]{fontenc}
\usepackage{natbib}
\usepackage[most]{tcolorbox}
\usepackage{multirow}

\usepackage[caption=false,justification=justified]{subfig}
\begin{document}
\title{Nano-mirror induced three-level quantum heat engine}
\author{Rejjak Laskar\footnote{Email: laskarrejjak786@gmail.com}}
\affiliation
{\it
Department of Physics, Aliah University, IIA/27, Newtown, Kolkata 700160, India
}

\date{\today}
\begin{abstract}
We propose a theoretical model that integrates a three-level $\Lambda$-type quantum heat engine with a vibrating nanomirror, where the connection is established via a laser field. In the presence of both hot and cold thermal photonic baths, the atom interacts with the laser field, generating photons as output, mimicking the operation of a heat engine driven by nanomirror vibrations. Using a semiclassical approach, we observe that the classical output or gain of the quantum heat engine is maximized as the photon distribution in the baths increases, provided that the coupling strength between the nanomirror and the engine is minimized. The model suggests that a greater temperature difference between the hot and cold reservoirs results in a more effective positive gain in the output. Thermodynamic analysis of the proposed model indicates that the total energy absorbed by the atomic system is equal to the energy released by the system, thus satisfying the first law of thermodynamics. The efficiency of the proposed engine decreases with an increasing photon distribution number in the hot reservoir, with a more pronounced decrease observed at higher values of atom-mirror coupling strength.
\end{abstract}
\maketitle
 \section{Introduction}
 
 A heat engine is a device that converts thermal energy into mechanical work by absorbing heat from a hot reservoir and expelling it into a cold one \cite{zemansky_heat_1981}. In the current model, we focus on quantum heat engines, which integrate principles from both quantum mechanics and classical thermodynamics \cite{doi:10.1146/annurev-physchem-040513-103724}. The study of quantum heat engines traces back to Scovil's work \cite{PhysRevLett.2.262}, where a three-level maser was demonstrated to function as a heat engine, constrained by Carnot efficiency \cite{PhysRev.156.343}. The primary goal of a three-level laser heat engine is to transform population inversion, a quantum phenomenon, into light. Over time, the three-level system has become a fundamental model in the study of quantum heat engines and refrigerators \cite{doi:10.1063/1.1776991, PhysRevE.49.3903, doi:10.1063/1.471453, doi:10.1080/09500340110090477, PhysRevLett.87.220601, HUMPHREY2005390, PhysRevA.74.063823, PhysRevLett.98.240601, scully2011quantum, Harbola_2012, PhysRevA.88.013842, PhysRevX.5.031044, PhysRevA.94.053859, PhysRevA.96.063806, PhysRevLett.119.050602, PhysRevA.103.062205, PhysRevResearch.2.043187}.

The present work primarily builds on the study by Harris \cite{PhysRevA.94.053859}, where a quantum heat engine is developed using a three-level $\Lambda$-type system within the framework of electromagnetically induced transparency (EIT) \cite{PhysRevLett.64.1107, PhysRevLett.66.2593}. This approach generates more photons in the output than the number of photons in the reservoir, as predicted by Kirchhoff's law. This model was later experimentally demonstrated using cold $^{85}\textrm{Rb}$ atoms \cite{PhysRevLett.119.050602}. Harris's work treats the output field as a classical electromagnetic field, exploring how system parameters affect its brightness \cite{PhysRevA.94.053859}. Studying the quantization of the emitted field offers insight into quantum effects. Recently, a single-atom quantum heat engine model based on EIT has been developed \cite{PhysRevA.109.012207}.

The present study proposes a novel quantum heat engine using nanomechanical mirror vibrations without cavity confinement, unlike typical optomechanical systems. Laser-induced vibrations in an atomic laser setup cause phase modulation and sidebands  \cite{PhysRevLett.107.223001, PhysRevA.82.021803, PhysRevA.93.023816}. The engine leverages EIT’s sensitivity to probe fields, making it highly responsive to perturbations, offering a new approach for enhancing quantum heat engines in optomechanics. We address two key questions: how nanomirror coupling affects the engine's gain and efficiency, and whether the engine satisfies the first law of thermodynamics.

\section{Model and gain}\label{tool}
 \begin{figure*}[!th]
\begin{center}
\begin{tikzpicture}[
      scale=.3,
      level/.style={thick},
      virtual/.style={thick,densely dashed},
      trans/.style={thick,<->,shorten >=2pt,shorten <=2pt,>=stealth},
      photon/.style={thick,->,shorten >=0pt,decorate, decoration={snake}},
      classical/.style={thick,densely dashed,draw=blue}
       >=stealth',
  pos=.8,
  photon/.style={decorate,decoration={snake,post length=3mm}}
    ]   
    \draw [line width=0.5mm] [line width=0.8mm](24.5cm,-42.5em) -- (29.5cm,-42.5em) node[right] {$\vert \textrm{g}\rangle $};
    \draw [line width=0.5mm] [line width=0.8mm](42cm,-35em) -- (47cm,-35em) node[right] {$\vert \textrm{g}^{'}\rangle $};
    \draw [line width=0.5mm] [line width=0.8mm](30cm,-15em) -- (40cm,-15em) node[right] {$\vert e\rangle $};
 \draw [line width=2mm] [line width=2mm](19.7cm,20em) -- (19.7cm,2em) ;
 \draw [line width=0.5mm] [line width=0.5mm](19.8cm,-62em) -- (19.8cm,-66em) ;
\draw [line width=0.5mm] [line width=0.5mm](18cm,-58em) -- (18cm,-68em) ;

\node[] at (19.8cm,-57.5em) {$\texttt{z}_{0}$};
    \node[] at (66cm,-64.5em) {$\texttt{z}$};
    \node[] at (37cm,26.5em) {$\textcolor{blue}{k_c, \omega_{c}}$};

   \node[] at (42cm,-17.5em) {$+\omega_{\texttt{m}}$};
   \node[] at (42cm,-21.5em) {$-\omega_{\texttt{m}}$};
    \node[] at (60.5cm,-18em) {$L$};
    \node[] at (19.8cm,23em) {$\omega_{\texttt{m}}$};
    \node[] at (61cm,22.5em) {$\textcolor{red}{\texttt{T}_{h}}$};
    \node[] at (62cm,-8em) {$\textcolor{red}{\texttt{T}_{c}}$};

   \draw[trans] (38cm,-20em) -- (38cm,-15em) node[] {$ $};
\fill[gray!50] (61,3) ellipse (7.5 and 2);
  \draw[][>=latex,thick,black](56,2.6) -- (50,-4)node[]{};
  \draw[][>=latex,thick,black](22,-4) -- (56,2.6)node[]{};
     \draw[][>=latex,thick,black](22,-4) -- (50,-4)node[]{};
      \draw[][>=latex,thick,black](22,-4) -- (22,-20)node[]{};
      \draw[][>=latex,thick,black](22,-20) -- (50,-20)node[]{};
       \draw[][>=latex,thick,black](50,-4) -- (50,-20)node[]{};
\filldraw [black] (57cm,5em) circle (8pt);
\filldraw [black] (58cm,5em) circle (8pt);
\filldraw [black] (59cm,5em) circle (8pt);
\filldraw [black] (56cm,7em) circle (8pt);
\filldraw [black] (63cm,6em) circle (8pt);
\filldraw [black] (61cm,7em) circle (8pt);
\filldraw [black] (60cm,5em) circle (8pt);
\filldraw [black] (64cm,8em) circle (8pt);
\filldraw [black] (59cm,8em) circle (8pt);
\filldraw [black] (57cm,8em) circle (8pt);
\filldraw [black] (59cm,8em) circle (8pt);
\filldraw [black] (62cm,8em) circle (8pt);
\filldraw [black] (66cm,9em) circle (8pt);
\filldraw [black] (65cm,11em) circle (8pt);
\filldraw [black] (57cm,11em) circle (8pt);
\filldraw [black] (59cm,11em) circle (8pt);
\filldraw [black] (62cm,11em) circle (8pt);
\draw[virtual] (36cm,-21.5em) -- (40.5cm,-21.5em);
\draw[virtual] (36cm,-19.5em) -- (40.5cm,-19.5em);
\draw[virtual] (19.5cm,10em) -- (19.1cm,19.8em);
\draw[virtual] (20cm,10em) -- (20.4cm,19.8em);
   \path[draw=red,solid,line width=0.5mm,fill=red,
preaction={-stealth,very thick,draw,red,shorten >=-1mm}
] (33.5cm,-15em) -- (27.5cm,-42em) node[midway,left] {};
\path[draw=blue,solid,line width=.5mm,fill=blue,
preaction={stealth-stealth,very thick,draw,blue,shorten >=-1mm}
] (44.5cm,-34.75em) -- (38.5cm,-20em) node[midway,right] {$\textcolor{blue}{\Omega_{\texttt{c}}}$};
\path[draw=black,solid,line width=.5mm,fill=black,
preaction={stealth-stealth,very thick,draw,black,shorten >=-1mm}
] (66cm,-16em) -- (55cm,-16em) node[midway,right] {};
\path[draw=black,solid,line width=0.5mm,fill=black,
preaction={-latex,very thick,draw,black,shorten >=-1mm}
] (18cm,-65em) -- (65cm,-65em) node[] {};

\path[draw=blue,solid,line width=0.5mm,fill=blue,
preaction={latex-,very thick,draw,blue,shorten >=-1mm}
] (50cm,28em) -- (70cm,38em) node[] {};
\path[draw=blue,solid,line width=0.5mm,fill=blue,
preaction={latex-,very thick,draw,blue,shorten >=-1mm}
] (35cm,21em) -- (50cm,28em) node[] {};
\path[draw=blue,solid,line width=0.5mm,fill=blue,
preaction={latex-,very thick,draw,blue,shorten >=-1mm}
] (20cm,15em) -- (35cm,21em) node[] {};
\path[draw=blue,solid,line width=0.5mm,fill=blue,
preaction={latex-,very thick,draw,blue,shorten >=-1mm}
] (34.5cm,23em) -- (39.5cm,25em) ;

\path[draw=blue,solid,line width=0.5mm,fill=blue,
preaction={-latex,very thick,draw,blue,shorten >=-1mm}
] (35cm,12.5em) -- (50cm,10em) node[] {};
\path[draw=blue,solid,line width=0.5mm,fill=blue,
preaction={-latex,very thick,draw,blue,shorten >=-1mm}
] (20cm,15em) -- (35cm,12.5em) node[] {};

\path[draw=blue,solid,line width=0.5mm,fill=blue,
preaction={-latex,very thick,draw,blue,shorten >=-1mm}
] (50cm,10em) -- (70cm,5em) node[] {};

\path[draw=red,solid,line width=2mm,fill=blue,
preaction={-latex,very thick,draw,red,shorten >=-3.5mm}
] (57cm,20em) -- (57cm,15em) node[] {};
\path[draw=red,solid,line width=2mm,fill=blue,
preaction={-latex,very thick,draw,red,shorten >=-3.5mm}
] (59cm,20em) -- (59cm,15em) node[] {};
\path[draw=red,solid,line width=2mm,fill=blue,
preaction={-latex,very thick,draw,red,shorten >=-3.5mm}
] (61cm,20em) -- (61cm,15em) node[] {};
\path[draw=red,solid,line width=2mm,fill=blue,
preaction={-latex,very thick,draw,red,shorten >=-3.5mm}
] (63cm,20em) -- (63cm,15em) node[] {};
\path[draw=red,solid,line width=2mm,fill=blue,
preaction={-latex,very thick,draw,red,shorten >=-3.5mm}
] (65cm,20em) -- (65cm,15em) node[] {};
\path[draw=red,solid,line width=2mm,fill=blue,
preaction={-latex,very thick,draw,red,shorten >=-4.3mm}
] (63cm,-5em) -- (63cm,0em) node[] {};
\path[draw=red,solid,line width=2mm,fill=blue,
preaction={-latex,very thick,draw,red,shorten >=-4.3mm}
] (61cm,-5em) -- (61cm,0em) node[] {};
\path[draw=red,solid,line width=2mm,fill=blue,
preaction={-latex,very thick,draw,red,shorten >=-4.3mm}
] (59cm,-5em) -- (59cm,0em) node[] {};
\path[draw=red,solid,line width=2mm,fill=blue,
preaction={-latex,very thick,draw,red,shorten >=-4.3mm}
] (57cm,-5em) -- (57cm,0em) node[] {};

\path[draw=red,solid,line width=2mm,fill=blue,
preaction={-latex,very thick,draw,red,shorten >=-4.3mm}
] (65cm,-5em) -- (65cm,0em) node[] {};
\end{tikzpicture}
\end{center}
\caption{ An ensemble of atoms (dots) is coupled by a mechanically oscillating mirror via control (blue color line with wave number $k_c$, frequency $\omega_{\textrm{c}}$) laser field.  The control laser field is reflected from the mirror before passing through the atomic system. The mirror vibrates with the frequency $\omega_{\textrm{m}}$ around its mean position $\textrm{z}_{0}$. Atoms are pumped by two black body radiation at temperatures $\textrm{T}_{h}$ and $\textrm{T}_{c}$. (Inset) The energy level diagram of the proposed three-level $\Lambda$-type heat engine, realized by control laser field of Rabi frequency $\Omega_{\textrm{c}}$. The control laser drives the transition $\ket{\textrm{g}^{'}}\leftrightarrow\ket{e}$. The black body radiation of temperatures $\textrm{T}_{h}$ and $\textrm{T}_{c}$ are pumping the transitions $\ket{\textrm{g}}\leftrightarrow\ket{e}$ and $\ket{\textrm{g}^{'}}\leftrightarrow\ket{e}$, respectively. Here, the atomic cell length is $L$.\label{HE}}
\end{figure*}
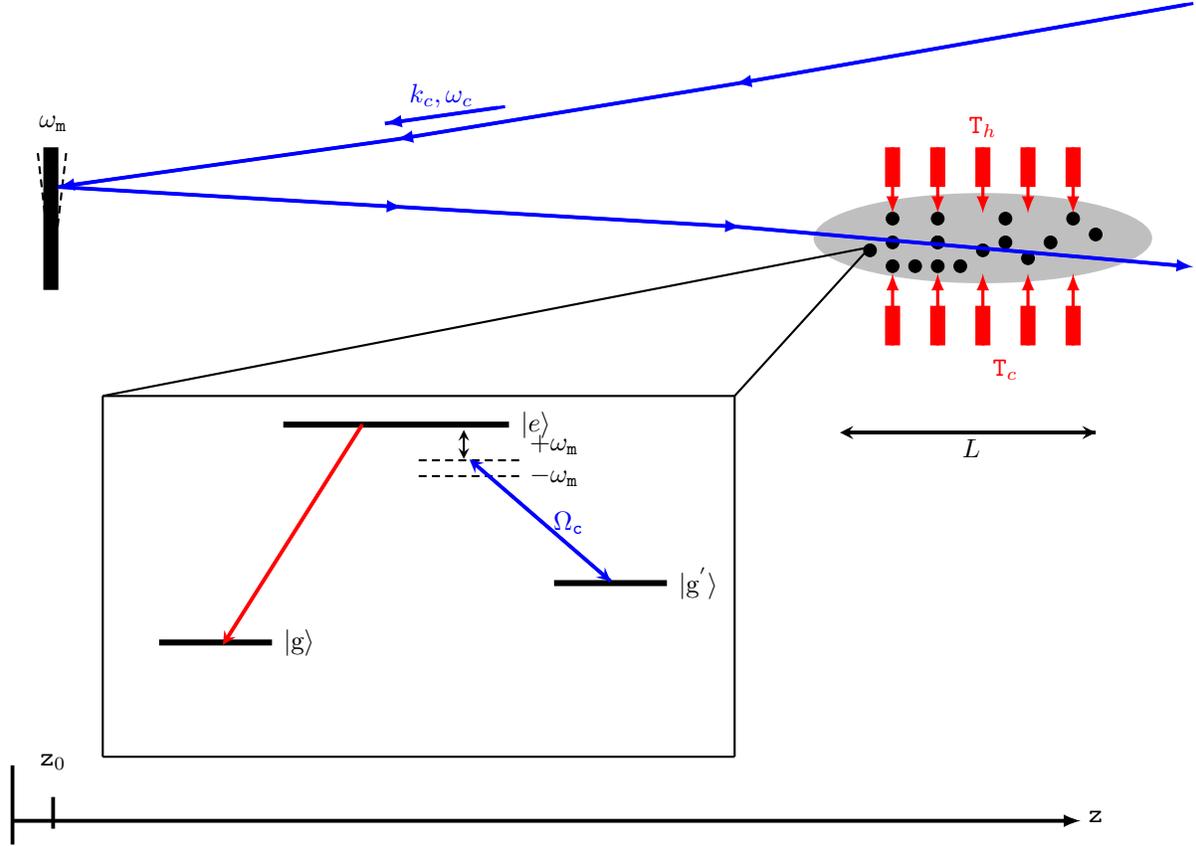
Figure \ref{HE} illustrates the schematic of a three-level $\Lambda$-type system in contact with two blackbody reservoirs at temperatures $\textrm{T}_{h}$ and $\textrm{T}_{c}$, as well as a coherent control laser field characterized by wave number $k_{c}$, frequency $\omega_{c}$, and Rabi frequency of $\Omega_{c}$. The control laser field (wave number $k_{c}$, frequency $\omega_{c}$) is reflected off a mirror of mass $\textrm{M}$ before passing through the atomic ensemble, shown by the blue line in the figure. The control field transmits the mirror's vibrations into the system. The mirror oscillates around its equilibrium position at $\textrm{z}=\textrm{z}_{0}$ with a frequency $\omega_{\textrm{m}}$. Since the control field is reflected off the vibrating mirror before reaching the atoms, the Rabi frequency becomes time-dependent, $\Omega_{\textrm{c}}(t)=\Omega_{\textrm{c}}\textrm{exp}(ik_{c}\textrm{z}_{\textrm{m}}(t))\approx  \Omega_{\textrm{c}}[1+ik_{c}\textrm{z}_{\textrm{m}}(t)]$.  Given that $k_{c}\textrm{z}_{\textrm{m}}(t)$ is small (i.e., the control field wavelength $\lambda_{\textrm{c}}$ is much larger than the mirror displacement $\textrm{z}_{\textrm{m}}(t)$), the control field's frequency shifts to $\omega_{c}\pm\omega_{\textrm{m}}$ due to the mirror's harmonic motion, $\textrm{z}_{\textrm{m}}(t)=\textrm{z}_{0}\cos(\omega_{\textrm{m}}t)$. The transitions $\ket{\textrm{g}}\rightarrow\ket{e}$ and $\ket{\textrm{g}^{'}}\rightarrow\ket{e}$ are driven by two blackbody reservoirs at temperatures $\textrm{T}_{h}$ and $\textrm{T}_{c}$, respectively. The control field couples the electric dipole (E1) allowed $\ket{\textrm{g}^{'}}\rightarrow\ket{e}$ transition. The two ground states $\ket{\textrm{g}}$ and $\ket{\textrm{g}^{'}}$ are metastable states and E1 forbidden. The engine operates through a series of transitions represented by $\ket{\textrm{g}} \xrightarrow{\textrm{T}_{h}}\ket{e}\xrightarrow{\textrm{T}_{c}}\ket{\textrm{g}^{'}}\xrightarrow{\omega_{\textrm{c}}}\ket{e}\xrightarrow{\omega_{\textrm{pr}}}\ket{\textrm{g}}$ \cite{PhysRevA.94.053859}. Here, $\omega_{\textrm{pr}}$ is the frequency of the emitted probe field.

The atomic Hamiltonian is defined as,
\begin{eqnarray}
\mathcal{H}_{0}=\hbar\sum_{i}\omega_{i}\ket{i}\bra{i};~~i=\textrm{g},~e,~\textrm{g}^{'}\end{eqnarray}
The interaction Hamiltonian of the system under both dipole- and rotating-wave approximations can be written as
 \begin{eqnarray}
\mathcal{H}_{\textrm{int}}=-\hbar\left[\Omega_{c}\hat{\sigma}_{e\textrm{g}^{'}}+\Omega_{c}^{*}\hat{\sigma}_{\textrm{g}^{'}e}+g_{\textrm{pr}}\hat{a}\hat{\sigma}_{e\textrm{g}}+g_{\textrm{pr}}\hat{a}^{\dagger}\hat{\sigma}_{\textrm{g}e}\right]
 \end{eqnarray}
Here, the $\hat{\sigma}_{jk}=\ket{j}\bra{k}$ is the transition operator, whereas the $\hat{a}$ and $g_{\textrm{pr}}$ are the anhilitian operator and vacuum Rabi frequency of the emitted probe field, respectively. The dynamics of the system is described by \cite{PhysRevA.109.012207}
  \begin{eqnarray}\label{OBE}
\begin{aligned}
\partial_{t}\rho =\frac{i}{\hbar}[\rho,\mathcal{H}]+\mathcal{L}_{h}(\rho)+\mathcal{L}_{c}(\rho)+\mathcal{L}_{cav}(\rho)
\end{aligned}
\end{eqnarray}
where, the total Hamiltonian is considered as $\mathcal{H}=\mathcal{H}_{0}+\mathcal{H}_{\textrm{int}}$. The 
$\mathcal{L}_{h}$ and $\mathcal{L}_{c}(\rho)$ are the Lindblad super operators for hot and cold reservoirs and given by
 \begin{eqnarray}\label{Lind}
 \begin{aligned}
     &\mathcal{L}_{h}(\rho)=\Gamma_{e\textrm{g}}({n}_{h}+1)\left(\hat{\sigma}_{\textrm{g}e}\rho\hat{\sigma}_{e\textrm{g}}-\frac{1}{2}\lbrace \hat{\sigma}_{ee},\rho\rbrace\right)\\&+\Gamma_{e\textrm{g}}{n}_{h}\left(\hat{\sigma}_{e\textrm{g}}\rho\hat{\sigma}_{\textrm{g}e}-\frac{1}{2}\lbrace\hat{\sigma}_{\textrm{g}\textrm{g}},\rho\rbrace\right),\\
     &\mathcal{L}_{c}(\rho)=\Gamma_{e\textrm{g}^{'}}({n}_{c}+1)\left(\hat{\sigma}_{\textrm{g}^{'}e}\rho\hat{\sigma}_{e\textrm{g}^{'}}-\frac{1}{2}\lbrace \hat{\sigma}_{ee},\rho\rbrace\right)\\&+\Gamma_{e\textrm{g}^{'}}{n}_{c}\left(\hat{\sigma}_{e\textrm{g}^{'}}\rho\hat{\sigma}_{\textrm{g}^{'}e}-\frac{1}{2}\lbrace \hat{\sigma}_{\textrm{g}^{'}\textrm{g}^{'}},\rho\rbrace\right),
     \end{aligned}
\end{eqnarray}
The rate at which emitted probe photons leak from the nanomirror cavity is given by $\kappa$ and corresponding Lindblad term is given by
\begin{eqnarray}
 \begin{aligned}
     &\mathcal{L}_{cav}(\rho)=\kappa\left(\hat{a}\rho\hat{a}^{\dagger}-\frac{1}{2}\hat{a}^{\dagger}\hat{a}\rho-\frac{1}{2}\rho\hat{a}^{\dagger}\hat{a}\right). 
 \end{aligned}
\end{eqnarray}
  $\Gamma_{jk}$ is the spontaneous decay rates for the transition $\ket{j}\leftrightarrow\ket{k}$. The respective photon distribution functions ${n}_{h}$ and ${n}_{c}$ for $\textrm{T}_{h}$ and $\textrm{T}_{c}$ reservoirs are defined as,
\begin{eqnarray}
\begin{aligned}
    &{n}_{h}=\frac{1}{\textrm{e}^{\left(\frac{\hbar\omega_{e\textrm{g}}}{\textrm{K}_{\textrm{B}}\textrm{T}_{h}}\right)}-1}\\
    &{n}_{c}=\frac{1}{\textrm{e}^{\left(\frac{\hbar\omega_{e\textrm{g}^{'}}}{\textrm{K}_{\textrm{B}}\textrm{T}_{c}}\right)}-1}
\end{aligned}
\end{eqnarray}
where $\hbar=\frac{h}{2\pi}$; $h$ and $\textrm{k}_{\textrm{B}}$ are the Planck's constant and the Boltzmann constant, respectively. $\omega_{jk}$ is the characteristic frequency corresponding to the transition $\ket{k}\rightarrow\ket{j}$. From equation \eqref{Lind}, the equation of motions are
\begin{scriptsize}
    \begin{eqnarray}\label{eqs}
    \begin{aligned}
&\partial_{t}\Bar{\rho}_{\textrm{gg}}=-\Gamma_{e\textrm{g}}n_{h}{\Bar{\rho}_{\textrm{gg}}}+\Gamma_{e\textrm{g}}(n_{h}+1){\Bar{\rho}_{ee}}+ig_{\textrm{pr}}\left({\hat{a}^{\dagger}\Bar{\rho}_{\textrm{g}e}}-{\hat{a}\Bar{\rho}_{e\textrm{g}}}\right)\\
         &\partial_{t}{\Bar{\rho}_{\textrm{g}^{'}\textrm{g}^{'}}}=-\Gamma_{e\textrm{g}^{'}}n_{c}{\Bar{\rho}_{\textrm{g}^{'}\textrm{g}^{'}}}+\Gamma_{e\textrm{g}^{'}}(n_{c}+1){\Bar{\rho}_{ee}}-i\Omega_{c}{\Bar{\rho}_{e\textrm{g}^{'}}}+i\Omega_{c}{\Bar{\rho}_{\textrm{g}^{'}e}}\\
&\partial_{t}{\Bar{\rho}_{\textrm{g}\textrm{g}^{'}}}=-\gamma_{\textrm{g}\textrm{g}^{'}}{\Bar{\rho}_{\textrm{g}\textrm{g}^{'}}}+i\Omega_{c}^{*}{\Bar{\rho}_{\textrm{g}e}}-ig_{\textrm{pr}}{\Bar{\rho}_{e\textrm{g}^{'}}\hat{a}}+\hat{F}_{\textrm{g}\textrm{g}^{'}}\\
&\partial_{t}{\Bar{\rho}_{\textrm{g}e}}=-\gamma_{\textrm{g}e}{\Bar{\rho}_{\textrm{g}e}}+i\Omega_{c}{\Bar{\rho}_{\textrm{g}\textrm{g}^{'}}}-ig_{\textrm{pr}}{(\Bar{\rho}_{ee}-\Bar{\rho}_{\textrm{g}\textrm{g}})\hat{a}}+\hat{F}_{\textrm{g}e}\\
&\partial_{t}{\Bar{\rho}_{\textrm{g}^{'}e}}=-\gamma_{\textrm{g}^{'}e}{\Bar{\rho}_{\textrm{g}^{'}e}}-i\Omega_{c}({\Bar{\rho}_{ee}-\Bar{\rho}_{\textrm{g}^{'}\textrm{g}^{'}}})+ig_{\textrm{pr}}{\Bar{\rho}_{\textrm{g}^{'}\textrm{g}}\hat{a}}+\hat{F}_{\textrm{g}^{'}e}
          \end{aligned}
       \end{eqnarray}
\end{scriptsize}

here the dephasing rates ($\gamma_{jk}$) corresponding to  the density matrix (${\rho}_{jk}$) are obtained as \cite{PhysRevA.109.012207},
\begin{eqnarray}
    \begin{aligned}
      &\gamma_{e\textrm{g}}=\frac{\Gamma_{e\textrm{g}}({n}_{h}+1)+\Gamma_{e\textrm{g}}{n}_{h}+\Gamma_{e\textrm{g}^{'}}({n}_{c}+1)}{2}\\
      &\gamma_{e\textrm{g}^{'}}=\frac{\Gamma_{e\textrm{g}^{'}}({n}_{c}+1)+\Gamma_{e\textrm{g}^{'}}{n}_{c}+\Gamma_{e\textrm{g}}({n}_{h}+1)}{2}\\
       &\gamma_{gg^{'}}=\frac{\Gamma_{e\textrm{g}}{n}_{h}+\Gamma_{e\textrm{g}^{'}}{n}_{c}}{2}
    \end{aligned}
\end{eqnarray}
The equation of motion of the annihilation operator ($\hat{a}$) can be written as \cite{PhysRevA.109.012207},
  \begin{eqnarray}\label{eqs}
\partial_{t}{\hat{a}}=-\frac{\kappa}{2}{\hat{a}}+ig_{\textrm{pr}}{\rho_{\textrm{g}e}}
\end{eqnarray}
 
The density matrix elements are periodic functions because of the coupling of the harmonic motion of the nanomirror with the atomic medium via the control field. Therefore, the density matrix elements can be expanded in a Fourier series with a periodicity of $\frac{2\pi}{\omega_{\textrm{m}}}$ (mirror's frequency $\omega_{\textrm{m}}$) as follows:
  \begin{eqnarray}\label{series}
    \Tilde{\rho}_{jk}=\sum_{l=-\infty}^{\infty}\Tilde{\rho}_{jk, l}\textrm{exp}(-il\omega_{\textrm{m}}t)
\end{eqnarray}
The amplitudes of the Fourier series \textit{i.e.} $\Tilde{\rho}_{jk, l}$ in the steady state satisfy,
\begin{eqnarray}\label{steady}
\Dot{\Tilde{\rho}}_{jk, l}=0;~~\textrm{All the terms are ignored for }  \vert l\vert>1 
\end{eqnarray}
where the constant density matrix elements $\Tilde{\rho}_{jk, 0}$ (the usual solution when nanomirror vibration is missing) and the first harmonics in nanomirror frequency $\Tilde{\rho}_{jk, \pm}$ are retained in the Fourier expansion [\eqref{series}] as,
\begin{eqnarray}\label{density}
    \Tilde{\rho}_{jk}=\Tilde{\rho}_{jk, 0}+\Tilde{\rho}_{jk, +}\textrm{exp}(-i\omega_{\textrm{m}}t)+\Tilde{\rho}_{jk, -}\textrm{exp}(i\omega_{\textrm{m}}t).
\end{eqnarray}
The steady state coherence term with Harmonic nanomirror vibration in interaction picture is obtained as,
\begin{scriptsize}
    \begin{eqnarray}\label{coherence}
    \begin{aligned}
       \Tilde{\rho}_{e\textrm{g}}&=-\frac{-\frac{1}{2}i(\rho_{ee}-\rho_{\textrm{g}\textrm{g}})(\frac{\gamma_{\textrm{g}\textrm{g}^{'}}}{2}\pm i\omega_{\textrm{m}})g_{\textrm{pr}}\hat{a}+\frac{i(1+\frac{\eta}{2}(\rho_{ee}-\rho_{\textrm{g}^{'}\textrm{g}^{'}}))g_{\textrm{pr}}\hat{a}\Omega_{\textrm{c}}^{2}}{4(\gamma_{e\textrm{g}^{'}}\pm2i\omega_{\textrm{m}})}}{-\frac{\gamma_{e\textrm{g}}}{2}(\frac{\gamma_{\textrm{g}\textrm{g}^{'}}}{2}\pm i\omega_{\textrm{m}})-\frac{1}{4}(1+\frac{\eta}{2})\Omega_{\textrm{c}}^{2}}
    \end{aligned}
\end{eqnarray}  
\end{scriptsize}

Similarly, the population terms are obtained as,
 \begin{eqnarray}
  \begin{aligned}
    \rho_{\textrm{g}\textrm{g}}&=\frac{X(\Gamma_{e\textrm{g}}+\textrm{R}_{h})}{3X\textrm{R}_{h}+X\Gamma_{e\textrm{g}}+\textrm{R}_{h}\Gamma_{e\textrm{g}^{'}}}  \\
     \rho_{ee}&=\frac{\textrm{R}_{h}X}{3X\textrm{R}_{h}+X\Gamma_{e\textrm{g}}+\textrm{R}_{h}\Gamma_{e\textrm{g}^{'}}}\\
       \rho_{\textrm{g}^{'}\textrm{g}^{'}}&=-\frac{-X\textrm{R}_{h}-\textrm{R}_{h}\Gamma_{e\textrm{g}^{'}}}{3X\textrm{R}_{h}+X\Gamma_{e\textrm{g}}+\textrm{R}_{h}\Gamma_{e\textrm{g}^{'}}}
 \end{aligned}
 \end{eqnarray}
 where,
 \begin{eqnarray}
  \begin{aligned}
    &X=\textrm{R}_{c}+\frac{\gamma_{e\textrm{g}^{'}}(1+\frac{\eta}{2})\Omega_{\textrm{c}}^{2}}{\gamma_{e\textrm{g}^{'}}^{2}+4\omega_{\textrm{m}}^{2}}\\
    &\eta=\textrm{k}_{c}\textrm{z}_{0},~\textrm{R}_{h}=\Gamma_{e\textrm{g}}\Bar{n}_{h}~\textrm{and}~\textrm{R}_{c}=\Gamma_{e\textrm{g}^{'}}\Bar{n}_{c}
 \end{aligned}
 \end{eqnarray}
The atom-nanomirror coupling strength is denoted by $\eta$. 

For analysis, we have implemented the model in the D$_{1}$-line of the $^{87}$Rb atomic system. The levels $\ket{\textrm{g}}$ and $\ket{\textrm{g}^{'}}$ are assigned to the $5S_{\frac{1}{2}}$ level with $F=1$, $m_{F}=0$ and $F=2$, $m_{F}=0$, respectively. The level $\ket{e}$ is equivalent to $5P_{\frac{1}{2}}$ with $F=2$, $m_{F}=-1$. For D$_{1}$-line, the natural decay rates are $\Gamma_{e\textrm{g}}=\Gamma_{e\textrm{g}^{'}}=5.7$ MHz \cite{steck2001rubidium}. From \eqref{eqs} and \eqref{coherence}, we can write the rate of change of $\hat{a}$ as,
\begin{eqnarray}
    \partial_{t}{\hat{a}}=\mathcal{G}{\hat{a}}
\end{eqnarray}
where, 
\begin{scriptsize}
    \begin{eqnarray}
    \mathcal{G}=-\frac{\kappa}{2}-\frac{\frac{1}{2}(\rho_{ee}-\rho_{\textrm{g}\textrm{g}})(\frac{\gamma_{\textrm{g}\textrm{g}^{'}}}{2}\pm i\omega_{\textrm{m}})g_{\textrm{pr}}^{2}+\frac{-(1+\frac{\eta}{2}(\rho_{ee}-\rho_{\textrm{g}^{'}\textrm{g}^{'}}))g_{\textrm{pr}}^{2}\Omega_{\textrm{c}}^{2}}{4(\gamma_{e\textrm{g}^{'}}\pm2i\omega_{\textrm{m}})}}{-\frac{\gamma_{e\textrm{g}}}{2}(\frac{\gamma_{\textrm{g}\textrm{g}^{'}}}{2}\pm i\omega_{\textrm{m}})-\frac{1}{4}(1+\frac{\eta}{2})\Omega_{\textrm{c}}^{2}}
\end{eqnarray}
\end{scriptsize}
       \begin{figure*}
\begin{center}
\end{center}
\begin{center}
\begin{tabular}{cc}
\centering
\textbf{(a)} &\textbf{(b)}\\
\includegraphics[scale=0.5]{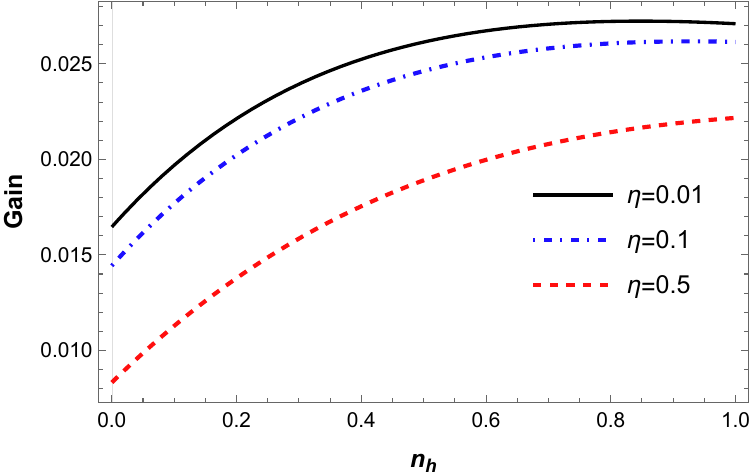}& \includegraphics[scale=0.5]{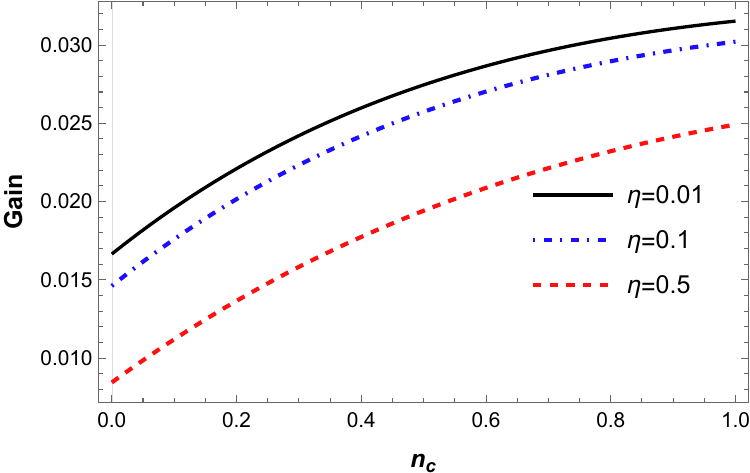}\\
\end{tabular}
\caption{Plots of the gain in the output of quantum heat engine are shown against photon distribution number (a) $n_{h}$ and (b) $n_{c}$ for various values of atom-nanomirror coupling strength $\eta$. In both plots (a) and (b), the control field strength is set to $\Omega_{\textrm{c}}=10$ MHz, and the mirror's frequency is $\omega_{m}=2$ MHz. The solid black, dot-dashed blue, and dashed red curves represent atom-nanomirror coupling strengths of $\eta=0.01$, $\eta=0.1$ and $\eta=0.5$, respectively. \label{gain_eta} }  
\end{center}
\end{figure*}
    
In our proposed quantum heat engine model (figure \ref{HE}), $\mathcal{G}$ can be interpreted as the output rate of the emitted probe field or gain from the engine. Figures \ref{gain_eta}(a) and \ref{gain_eta}(b) illustrate that the gain increases with the photon distribution numbers $n_{h}$ and $n_{c}$ of the respective hot and cold reservoirs. Figures \ref{gain_eta}(a) and \ref{gain_eta}(b) also clearly show that as the coupling strength ($\eta$) between the atom and the nanomirror increases, the gain decreases. This is understandable given that the perturbation induced by the nanomirror affects the atomic system and, consequently, influences the output rate of emitted probe photons.
   \begin{figure*}
\begin{center}
\end{center}
\begin{center}
\begin{tabular}{cc}
\centering
\textbf{(a)} &\textbf{(b)}\\
\includegraphics[scale=0.5]{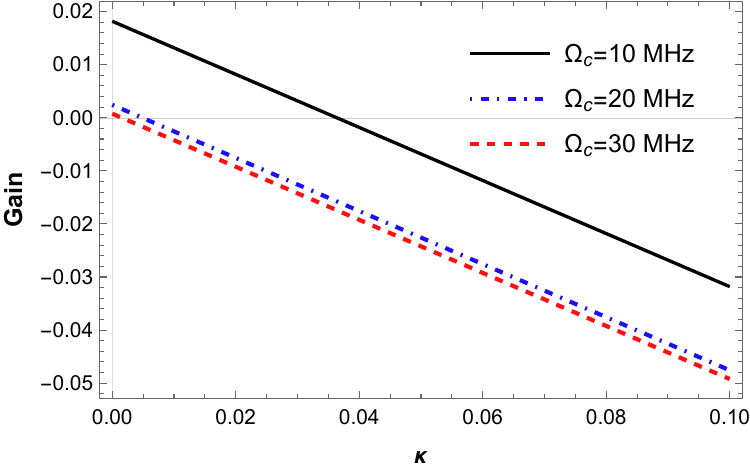}& \includegraphics[scale=0.5]{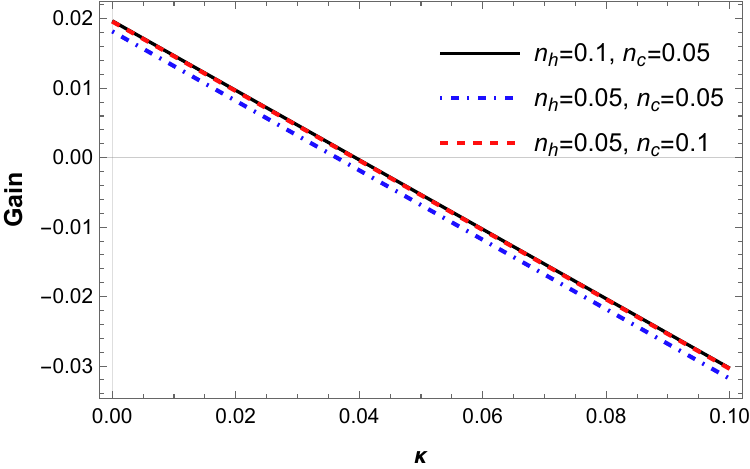}\\
\end{tabular}
\caption{Plots of the gain in the output of quantum heat engine are shown against emitted probe photon leaking rate $\kappa$ for different strength of control field strength (a) $\Omega_{\textrm{c}}$ and (b) photon distribution numbers $n_{h}$ and $n_{c}$. In plot (a), the photon distribution numbers are $n_{h}=0.05$ and $n_{c}=0.05$ and in plot (b), the control field strength is $\Omega_{\textrm{c}}=10$ MHz. The mirror's frequency is considered as $\omega_{m}=2$ MHz. The solid black, dot-dashed blue, and dashed red curves represent (a) $\Omega_{\textrm{c}}$=10 MHz, $\Omega_{\textrm{c}}$=20 MHz and $\Omega_{\textrm{c}}$=30 MHz, respectively. The solid black, dot-dashed blue, and dashed red curves represent (b) ($n_{h}=0.1$, $n_{c}=0.05$),  ($n_{h}=0.05$, $n_{c}=0.05$) and ($n_{h}=0.05$, $n_{c}=0.1$), respectively. \label{gain_kappa} }  
\end{center}
\end{figure*}
 Figure \ref{gain_kappa} shows that as the leakage rate of the emitted probe photons ($\kappa$) increases, the gain becomes negative, which is expected. However, the interesting observation is that with increasing control field strength, the gain becomes negative more rapidly with respect to $\kappa$ [figure \ref{gain_kappa} (a)]. This indicates that in a quantum heat engine, a stronger control field makes the engine's output more sensitive to the leakage rate. On the other hand, when the photon distribution numbers in both the hot and cold reservoirs are the same, the gain tends to become more negative as $\kappa$ increases, as shown in figure \ref{gain_kappa}(b). This suggests that the temperatures of the two reservoirs should not be equal as the photon distribution increases with temperature. Therefore, a temperature difference is necessary to maintain positive gain.






\begin{figure}
    \centering
    \includegraphics[width=0.7\linewidth]{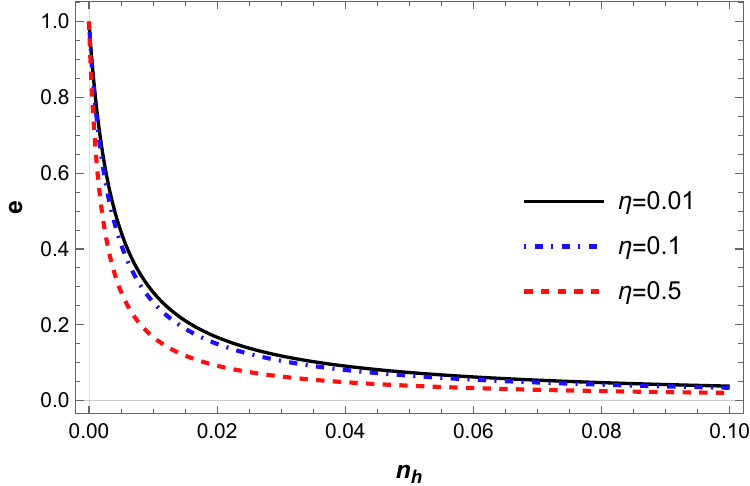}
    \caption{The plots display the efficiency (e) as a function of the photon distribution number $n_{h}$ of the hot reservoir for different values of the atom-nanomirror coupling strength $\eta$. The control field strength is $\Omega_{\textrm{c}}=10$ MHz, and the mirror's frequency is $\omega_{m}=2$ MHz. The solid black, dot-dashed blue, and dashed red curves represent atom-nanomirror coupling strengths of $\eta=0.01$, $\eta=0.1$, and $\eta=0.5$, respectively.}
    \label{effeciency}
\end{figure}
\section{Thermodynamic analysis of quantum engine}
The energy of the unperturbed atomic system is given by $E=\textrm{Tr}(\Bar{\rho}\mathcal{H}_{0})$, and the first-order derivative of $E$ with respect to time, using \eqref{OBE}, is as follows \cite{PhysRevA.74.063823}
 \begin{eqnarray}\label{energyflux}
     \Dot{E}= \textrm{Tr}\lbrace\Dot{\Bar{\rho}}\mathcal{H}_{0}\rbrace=P_{c}-\Dot{Q}_{out}+\Dot{Q}_{h}+\Dot{Q}_{c}
 \end{eqnarray}
The energy absorbed per second by the atomic system from the control field is denoted as $P_{c}$, while the energy transferred from the atomic system to the cold reservoir is $\Dot{Q}_{c}$. These quantities are related by the expression $P_{c}+\Dot{Q}_{c}=-\textrm{Tr}\lbrace  \frac{i}{\hbar}[\mathcal{H}_{0},\mathcal{H}_{int}]\Bar{\rho}\rbrace$, where 
\begin{eqnarray}\label{power1}
P_{c}=i\hbar\omega_{e\textrm{g}^{'}}\Omega_{\textrm{c}}(\Bar{\rho}_{e\textrm{g}^{'}}-\Bar{\rho}_{\textrm{g}^{'}e})
    \end{eqnarray}
and
\begin{eqnarray}\label{power2}
    \Dot{Q}_{c}=\textrm{Tr}(\mathcal{L}_{c}(\Bar{\rho})\mathcal{H}_{0})=\hbar\omega_{e\textrm{g}^{'}}\Gamma_{e\textrm{g}^{'}}\left[(n_{c}+1)\Bar{\rho}_{ee}-n_{c}\Bar{\rho}_{\textrm{g}^{'}\textrm{g}^{'}})\right]
    \end{eqnarray}
The power of engine's output ($\Dot{Q}_{{out}}$) and rate of energy provided by the hot reservoir  ($\Dot{Q}_{h}$) are obtained as,
\begin{eqnarray}\label{power3}
\begin{aligned}
\Dot{Q}_{out}=i\hbar\omega_{e\textrm{g}}g_{\textrm{pr}}(\Bar{\rho}_{\textrm{g}e}\hat{a}^{\dagger}-\Bar{\rho}_{e\textrm{g}}\hat{a})
\end{aligned}
\end{eqnarray}
and
\begin{eqnarray}\label{power4}
\Dot{Q}_{h}=\textrm{Tr}(\mathcal{L}_{h}(\Bar{\rho})\mathcal{H}_{0})=\hbar\omega_{e\textrm{g}}\Gamma_{e\textrm{g}}\left[n_{h}\Bar{\rho}_{\textrm{g}\textrm{g}}-(n_{h}+1)\Bar{\rho}_{ee})\right]
    \end{eqnarray}
Comparing the \eqref{power3}, \eqref{power3} and \eqref{eqs}, we can find that in steady state condition ($\partial_{t}\Bar{\rho}_{\textrm{gg}}=0$),
\begin{eqnarray}\label{balance1}
   \Dot{Q}_{out}= \Dot{Q}_{h}
\end{eqnarray}
The equation above suggests that the output energy of the engine is equivalent to the energy extracted from the hot reservoir. Similarly, from \eqref{power1}, \eqref{power2} and \eqref{eqs}, we find
\begin{eqnarray}
    P_{c}=\Dot{Q}_{c}
\end{eqnarray}
It suggests that the energy absorbed from the control field is transferred to the cold reservoir. From \eqref{power1} to \eqref{power4}, it is evident that the amount of energy absorbed by the atomic system is equal to the amount of energy released, which aligns with the First Law of Thermodynamics. The efficiency of the engine is defined as
\begin{eqnarray}
    \textrm{e}=\frac{\Dot{Q}_{out}}{\Dot{Q}_{h}+ P_{c}}
\end{eqnarray}
The efficiency decreases as the photon distribution number in the hot reservoir increases, as shown in figure \ref{effeciency}. This decline becomes more pronounced with the increase in atom-nanomirror coupling strength.

\section{Conclusions}
In conclusion, we present a theoretical model integrating a three-level $\Lambda$-type quantum heat engine with a vibrating nanomirror, connected through a laser field. In this setup, the atom interacts with both hot and cold thermal photonic baths, generating photons as output, thereby mimicking a nanomirror-driven heat engine. Our semiclassical analysis shows that the classical output or gain of the engine is maximized with increasing photon distribution in the baths, particularly when the coupling strength between the nanomirror and the engine is minimized. The model further demonstrates that a larger temperature difference between the hot and cold reservoirs enhances the positive gain. Thermodynamic analysis confirms that the system adheres to the first law of thermodynamics, as the total energy absorbed equals the energy released. However, the efficiency decreases with increasing photon distribution in the hot reservoir, especially at higher atom-mirror coupling strengths.

\section*{Acknowledgment}The author extends heartfelt gratitude to Dr. Jayanta Kr. Saha and Dr. Md. Mabud Hossain for granting the freedom to carry out this research project.

\newpage
\bibliographystyle{unsrt}
\bibliography{biblio}

\end{document}